\documentclass[twocolumn,prb,superscriptaddress]{revtex4}
\usepackage{bm,epsfig,subfigure,times,amsmath,dcolumn,amssymb}
\usepackage{graphicx,epstopdf}
\usepackage{threeparttable}
\newcommand{\htwo}{H$_\text{2}$}

\newcommand{\libh}{LiBH$_\text{4}$}
\newcommand{\cabh}{Ca(BH$_\text{4}$)$_\text{2}$}

\newcommand{\alh}{AlH$_\text{3}$}
\newcommand{\cah}{CaH$_\text{2}$}
\newcommand{\mgh}{MgH$_\text{2}$}
\newcommand{\tih}{TiH$_\text{2}$}
\newcommand{\vh}{VH$_\text{2}$}
\newcommand{\sch}{ScH$_\text{2}$}
\newcommand{\crh}{CrH$_\text{2}$}
\newcommand{\alb}{AlB$_\text{2}$}
\newcommand{\mgbfour}{MgB$_\text{4}$}
\newcommand{\mgbtwo}{MgB$_\text{2}$}
\newcommand{\tib}{TiB$_\text{2}$}
\newcommand{\vb}{VB$_\text{2}$}
\newcommand{\scb}{ScB$_\text{2}$}
\newcommand{\crb}{CrB$_\text{2}$}
\newcommand{\cab}{CaB$_\text{6}$}

\begin{document}

\title{Thermodynamic guidelines for the prediction of hydrogen storage reactions and their application to destabilized hydride mixtures}

\author{Donald J. Siegel}
\affiliation{Ford Motor Company, MD1170/RIC, Dearborn, MI 48121}
\author{C. Wolverton}
\altaffiliation[Present address: ]{Department of Materials Science and Engineering, Northwestern University, Evanston, IL 60208}
\affiliation{Ford Motor Company, MD1170/RIC, Dearborn, MI 48121}
\author{V. Ozoli\c{n}\v{s}}
\affiliation{Department of Materials Science and Engineering, University of California, Los Angeles, CA 90095}

\date{\today}

\begin{abstract}
  We propose a set of thermodynamic guidelines aimed at facilitating
  more robust screening of hydrogen storage reactions.  The utility of
  the guidelines is illustrated by reassessing the validity of
  reactions recently proposed in the literature, and through vetting a
  list of more than 20 candidate reactions based on destabilized
  \libh\ and \cabh\ borohydrides.  Our analysis reveals several new
  reactions having both favorable thermodynamics and relatively high
  hydrogen densities (ranging from 5-9 wt.\% \htwo\ \& 85-100 g
  \htwo/L), and demonstrates that chemical intuition alone is not
  sufficient to identify valid reaction pathways.
\end{abstract}

\maketitle

\section{Introduction}
The potential of emerging technologies such as fuel cells (FCs) and
photovoltaics for environmentally-benign power generation has sparked
renewed interest in the development of novel materials for
high-density energy storage.  For mobile applications such as in the
transportation sector, the demands placed upon energy storage media
are especially stringent,\cite{pinkerton04} as the leading candidates
to replace fossil-fuel-powered internal combustion engines
(ICEs)---proton exchange membrane FCs and hydrogen-powered ICEs
(\htwo-ICEs)---rely on \htwo\ as a fuel.  Although \htwo\ has about
three times the energy density of gasoline by weight, its volumetric
density, even when pressurized to 10,000~psi, is roughly six times
less than that of gasoline. Consequently, safe and efficient storage
of \htwo\ has been identified\cite{crabtree04} as one of the key
scientific obstacles to realizing a transition to \htwo-powered
vehicles.

Perhaps the most promising approach to achieving the high \htwo\
densities needed for mobile applications is via absorption in
solids.\cite{schlapbach01}  Metal hydrides such as LaNi$_5$H$_6$ have
long been known to reversibly store hydrogen at volumetric densities
surpassing that of liquid \htwo, but their considerable weight results
in gravimetric densities that are too low for lightweight
applications.\cite{sandrock99}  Accordingly, recent
efforts\cite{soulie02,zuttel03,nakamori06a,lodziana04,miwa06_cabh}
have increasingly focused on low-Z complex hydrides, such as metal
borohydrides, $M$(BH$_4$)$_n$, where $M$ represents a metallic cation,
as borohydrides have the potential to store large quantities of
hydrogen (up to 18.5 wt.\% in \libh).  Nevertheless, the
thermodynamics of \htwo-desorption from known borohydrides are
generally not compatible with the temperature-pressure conditions of
FC operation: for example, in \libh\ strong hydrogen-host bonds result
in desorption temperatures in excess of 300$^\circ$C.\cite{zuttel03}
Thus the suitability of \libh\ and other stable hydrides as practical
\htwo-storage media will depend upon the development of effective
destabilization schemes.

Building on earlier work by Reilly and Wiswall,\cite{reilly68} Vajo
\textit{et al.}\cite{vajo05} recently demonstrated that \libh\ can be
destabilized by mixing with \mgh.  In isolation, the decomposition of
these compounds proceeds according to:
\begin{subequations}
\label{pure_rxns}
\begin{eqnarray}
\text{LiBH}_4 &\rightarrow& \text{LiH} + \text{B} + \frac{3}{2} \text{H}_2, \label{libh}\\
\text{MgH}_2 &\rightarrow& \text{Mg} + \text{H}_2, \label{mgh}
\end{eqnarray}
\end{subequations}
yielding 13.6 and 7.6 wt.\% \htwo, respectively, at temperatures above
300$^\circ$C.  The high desorption temperatures are consistent with
the the relatively high enthalpies of desorption: $\Delta H \sim$ 67
(\libh) and $\sim$70 (\mgh) kJ/(mol \htwo).\cite{vajo05,manchester00}
By mixing \libh\ with \mgh, $\Delta H$ for the combined reaction can
be decreased below those of the isolated compounds due to the
exothermic formation enthalpy of MgB$_2$:
\begin{equation}
\label{destab}
\text{LiBH}_4 + \frac{1}{2}\text{MgH}_2 \rightarrow \text{LiH} + \frac{1}{2}\text{MgB}_2 + 2 \text{H}_2.
\end{equation}
That is, formation of the MgB$_2$ product \textit{stabilizes} the
dehydrogenated state in Eq.~\ref{destab} relative to that of
Eq.~\ref{pure_rxns}, thereby \textit{destabilizing} both \libh\ and
\mgh.  By adopting this strategy, measured isotherms for the \libh\ +
$\frac{1}{2}$\mgh\ mixture over 315--400$^\circ$C exhibited a 25
kJ/mol \htwo\ decrease in $\Delta H$ relative \libh\ alone, with an
approximately tenfold increase in equilibrium \htwo\
pressure.\cite{vajo05}  In addition, the hydride mixture was shown to
be reversible with a density of 8--10 wt.\% \htwo.\cite{vajo05}
Nevertheless, the extrapolated temperature of $T = 225^\circ$C at
which $P_{\text{H}_2} = 1$~bar is still too high for mobile applications, and
suggests that \textit{additional} destabilization is necessary.

The concept of thermodynamic destabilization appears to offer new
opportunities for accessing the high \htwo\ content of strongly-bound
hydrides.  However, the large number of known hydrides suggests that
experimentally testing all possible combinations of known compounds
would be impractical; thus a means for rapidly screening for
high-density \htwo-storage reactions with appropriate
thermodynamics\cite{alapati06} would be of great
value.\footnote{Experimental testing of hydrogen storage
  materials---many of which are air-sensitive---can be a slow,
  painstaking process. For example, an equilibrium measurement of the
  extent of hydrogen desorption/uptake at a single temperature may
  require several \textit{months} to complete in kinetically-hindered
  materials.\cite{asudik_private} In contrast, the first-principles
  thermodynamic calculations presented here (encompassing more than 20
  unique reactions) were completed in 2-3 weeks.} Towards these ends,
here we employ first-principles calculations to identify new
\htwo-storage reactions with favorable temperature-pressure
characteristics based on destabilizing \libh\ and
\cabh\cite{miwa06_cabh} by mixing with selected metal hydrides.  Our
goal is to determine whether additional destabilization of \libh\ and
\cabh---beyond that demonstrated\cite{vajo05} with \libh/\mgh---is
possible by exploiting the exothermic formation enthalpies of the
metal borides. We focus specifically on thermodynamic issues since
appropriate thermodynamics is a necessary condition for any viable
storage material, and thermodynamic properties are not easily altered.
While kinetics must also be considered, catalysts and novel synthesis
routes have been shown to be effective at improving reversibility and
the rates of \htwo\ uptake/release.\cite{bogdanovic97} By screening
through $\sim$20 distinct reactions, we identify four new destabilized
mixtures having favorable Gibbs free energies of desorption in
conjunction with high gravimetric (5--9 wt.\%) and volumetric (85--100
g \htwo/L) storage densities.  The predicted reactions present new
avenues for experimental investigation, and illustrate that compounds
with low gravimetric densities (i.e., transition metal hydrides) may
yield viable \htwo-storage solutions when mixed with lightweight
borohydrides.  An advantage of the present approach is that it relies
only on known compounds with established synthesis routes, in contrast
to other recent studies which have proposed \htwo-storage reactions
based on materials which have yet to be
synthesized.\cite{deng04,zhao05,yildirim05,lee06,sun06}

An additional distinguishing feature of this study is the development
of a set of thermodynamic guidelines aimed at facilitating more robust
predictions of hydrogen storage reactions.  The guidelines are used to
vet the present set of candiate reactions, and to illustrate how other
reactions recently reported in the literature\cite{alapati06} are
thermodynamically unrealistic.  In total, this exercise reveals some
of the common pitfalls that may arise when attempting to simply
``guess'' at reaction mechanisms.

\section{Methodology}
Our first-principles calculations were performed using a
planewave-projector augmented wave method
(\textsc{vasp})\cite{kresse96, blochl94a} based on the generalized
gradient approximation\cite{perdew92} to density functional theory.
All calculations employed a planewave cutoff energy of 400~eV, and
k-point sampling was performed on a dense grid with an energy
convergence of better than 1~meV per supercell.  Internal atomic
positions and external cell shape/volume were optimized to a tolerance
of better than 0.01~eV/\AA.  Thermodynamic functions were evaluated
within the harmonic approximation,\cite{wallace72} and normal-mode
vibrational frequencies were evaluated using the so-called direct method
on expanded supercells.\cite{wei92,wolverton04,siegel07,wolverton07}
Further information regarding the details and experimental validation
of our calculations can be found
elsewhere.\cite{siegel07,wolverton04,wolverton07}


Our search for high-density \htwo-storage reactions is based on a
series of candidate reactions that are analogous to Eq.~\ref{destab}:
\begin{equation}
\label{gen_eqn}
y\,A(\text{BH}_4)_n + M\text{H}_x \rightarrow y\,A\text{H}_n +M\text{B}_{yn} +  \frac{3yn + x}{2}\,\text{H}_2,
\end{equation}
where $A$ = Li or Ca [$n$ = 1 (2) for Li (Ca)], $M$ represents a
metallic element, and the coefficients $x$ and $y$ are selected based
on the stoichiometries of known hydrides $M$H$_x$ and borides
$M$B$_{yn}$.  To maximize gravimetric density we limit $M$ to
relatively light-weight elements near the top of the periodic table.
In the case of $A$ = Li, the enthalpy of Eq.~\ref{gen_eqn} per mol
\htwo\ can be expressed as:
\begin{equation}
\label{gen_eqn_enthalpy}
  \Delta H = \frac{2}{3y+x}\left[\frac{3y}{2}\Delta H^\text{LiBH$_4$} + \frac{x}{2}\Delta H^\text{$M$H$_x$} - \Delta H^\text{$M$B$_y$}\right]
\end{equation}
where $\Delta H^i$ are the desorption (formation) enthalpies of the
respective hydrides (borides) per mol \htwo\ ($M$).  Thus $\Delta H$
for the destabilized \libh\ reaction is simply an average of the
hydride desorption enthalpies, less the enthalpy of boride formation.

\section{Results} 
\begin{table*}
  \caption{\htwo\ densities and calculated thermodynamic quantities for candidate \htwo\ storage reactions.  Units are J/K/mol \htwo\ for $\Delta S_\text{vib}$ and kJ/mol \htwo\ for $\Delta E$ and $\Delta H$; column 7 refers to the temperature at which $P_{\text{H}_2} = 1$ bar. Reactions denoted with a $*$ will not proceed as written (see text).  The enthalpies of reactions 24--27 have been measured in prior experiments, and are included here (in parentheses) to validate the accuracy of our calculations.  For comparison, system-level targets for gravimetric and volumetric density are cited in the bottom row.\cite{doe_targets}}
\label{table}
\begin{ruledtabular}
\begin{tabular}{llrrrrrr}
  Rxn.& \multicolumn{1}{c}{Reaction} & Wt.\% & Vol.\ density & $\Delta E$ & $\Delta H^\text{T=300K}$& $T$, $P$=1 bar&$\Delta S_\text{vib}^\text{T=300K}$ \\
  No. & &(kg \htwo/kg) & (g \htwo/L)&  & & ($^\circ$C) & \\ \hline
  1$*$ &4\libh\ + 2\alh\ $\rightarrow$ 2\alb\ + 4LiH + 9\htwo         & 12.4&106&54.8&39.6& 83&   $-$18.4\\
  2 &2\libh\ + Al $\rightarrow$ \alb\ + 2LiH + 3\htwo                 & 8.6 &80 &77.0&57.9& 277&  $-$26.9\\
  3$*$ &4\libh\ + \mgh\  $\rightarrow$ \mgbfour\ + 4LiH + 7\htwo      & 12.4&9 5&68.2&51.8& 206&  $-$23.3\\
  4$*$&2\libh\ + Mg$\rightarrow$ \mgbtwo\ + 2LiH + 3\htwo             & 8.9 &76 &65.9&46.4& 170&  $-$29.4\\
  5 &2\libh\ + \tih $\rightarrow$ \tib\ + 2LiH + 4\htwo               & 8.6 &103&21.4&4.5 & &  $-$23.3\\
  6 &2\libh\ + \vh$\rightarrow$ \vb\ + 2LiH + 4\htwo                  & 8.4 &105&24.7&7.2 & $-$238&  $-$21.7\\
  7 &2\libh\ + \sch$\rightarrow$ \scb\ + 2LiH + 4\htwo                & 8.9 &99 &48.8&32.6& 26&  $-$21.4\\
  8$*$ &2\libh\ + \crh$\rightarrow$ \crb\ + 2LiH + 4\htwo             & 8.3 &109&33.9&16.4& $-$135&  $-$19.2\\
  9$*$ &2\libh\ + 2Fe$\rightarrow$ 2FeB + 2LiH + 3\htwo               & 3.9 &76 &32.7&12.8& $-$163&  $-$24.6\\
  10&2\libh\ + 4Fe$\rightarrow$ 2Fe$_\text{2}$B + 2LiH + 3\htwo        & 2.3 &65 &21.6&1.2 & &  $-$24.4\\\vspace{0.07in}
  11&2\libh\ + Cr$\rightarrow$ \crb\ +2LiH + 3\htwo                   & 6.3 &84 &50.9&31.7& 25&  $-$23.8\\
  12&\cabh$\rightarrow$ $\frac{2}{3}$\cah\ + $\frac{1}{3}$\cab\ + $\frac{10}{3}$\htwo & 9.6&107 & 57.1& 41.4 &88 & $-$16.0 \\
  13$*$&\cabh\ + \mgh$\rightarrow$ \cah\ + \mgbtwo + 4\htwo           & 8.4 &99 & 61.6 & 47.0 & 135&$-$16.2\\
  14$*$&2\cabh\ + \mgh$\rightarrow$ 2\cah\ + \mgbfour\ + 7\htwo       & 8.5 &98 & 63.6 & 47.9 & 147& $-$17.0 \\
  15$*$&\cabh\ + Mg$\rightarrow$ \cah\ + \mgbtwo\ + 3\htwo            & 6.4 &79 & 60.6 & 41.9 & 111& $-$22.0\\
  16$*$&\cabh\ + Al$\rightarrow$ \cah\ + \alb\ + 3\htwo               & 6.3 &83 & 71.7 & 53.4 & 200& $-$19.5\\
  17$*$&\cabh\ + \alh$\rightarrow$ \cah\ + \alb\ + $\frac{9}{2}$\htwo & 9.1 &109& 51.2 & 36.6 & 39&$-$13.5\\
  18&\cabh\ + \sch$\rightarrow$ \cah\ + \scb\ + 4\htwo                & 6.9 &102& 44.8 & 29.2 & $-$20 &$-$15.9\\
  19&\cabh\ + \tih$\rightarrow$ \cah\ + \tib\ + 4\htwo                & 6.7 &106& 17.4 & 1.1 &&$-$17.7 \\
  20&\cabh\ + \vh$\rightarrow$  \cah\ + \vb\ + 4\htwo                 & 6.6 &108& 20.8 & 3.8 && $-$16.2\\
21$*$&\cabh\ + \crh$\rightarrow$ \cah\ + \crb\ + 4\htwo             & 6.5 &113& 29.9 & 13.1 & $-$180&$-$13.6\\\vspace{0.07in}
22&\cabh\ + Cr $\rightarrow$ \cah\ + \crb\ + 3\htwo                 & 5.0 &86 & 45.6 &27.2 &$-$38 & $-$16.4\\
23&6\libh\ + \cah$\rightarrow$ \cab\ + 6LiH + 10\htwo               & 11.7&93 & 61.9 (63)\footnotemark[1]&45.4& 146&$-$22.7\\
24&2\libh\ + \mgh$\rightarrow$ \mgbtwo\ + 2LiH + 4\htwo             & 11.6&96 &65.6&50.4 (41)\footnotemark[2] & 186& $-$21.7\\
25&2\libh $\rightarrow$ 2LiH + 2B + 3 \htwo                         & 13.9&93 &81.4&62.8 (67)\footnotemark[2] & 322&$-$27.1\\
26&\libh $\rightarrow$ Li + B + 2\htwo                              & 18.5&124&103.5&89.7 (95)\footnotemark[3]& 485&$-$15.3\\
27&\mgh$\rightarrow$ Mg + \htwo                                     & 7.7 &109&64.5&62.3 &195 & 1.3\\
& & & & &  (65.8--75.2)\footnotemark[4] & \\
& \multicolumn{1}{c}{U.S. DOE system-level targets (2010/2015)}     & 6/9 &45/81& & & &  \\
\end{tabular}
\footnotetext[1]{Ref.~\onlinecite{alapati06}; $^\text{\textit{b}}$Ref.~\onlinecite{vajo05}; $^\text{\textit{c}}$Ref.~\onlinecite{janaf98}; $^\text{\textit{d}}$Ref.~\onlinecite{manchester00}}
\end{ruledtabular}
\end{table*}

Table~\ref{table} lists theoretical \htwo\ densities, and calculated
dehydrogenation enthalpies and entropies for several potential
\htwo-storage reactions.  Reactions 1--22 enumerate the candidate new
reactions, while reactions 23--27 are included in order to validate
the accuracy of our predictions by comparing with
experimentally-measured enthalpies\cite{vajo05,janaf98,manchester00}
and previous first-principles results\cite{alapati06} (shown in
parentheses). Turning first to the reactions from experiment (24--27),
it is clear that the calculated $T = 300$~K enthalpies are generally
in good agreement with the measured data.  As mentioned above,
reaction 24 was studied by Vajo and co-workers\cite{vajo05} (see
Eq.~\ref{destab}).  Our calculated enthalpy of 50.4 kJ/mol \htwo\
overestimates the experimental value by $\sim$10 kJ/mol.  However,
since the experimental measurements were made at temperatures ($T = $
315--400$^\circ$C) above the \libh\ melting point ($T_m = $
268$^\circ$C),\cite{zuttel03}) and our calculations are with respect to
the ground state \textit{Pnma} crystal structure,\cite{zuttel03} we
expect $\Delta H^\text{calc}\textit{(Pnma)} > \Delta
H^\text{expt}\textit{(liquid)}$ due to the higher enthalpy of the
liquid state.

We begin our discussion of the candidate reactions by commenting on
the vibrational contributions ($\Delta S_\text{vib}$) of the solid
state phases to the total dehydrogenation entropy, $\Delta S$.  Based
on the notion that $\Delta S$ is largely due to the entropy of \htwo\ 
[$\Delta S \simeq S_0^{H_2} \simeq 130 ~$J/(mol K) at 300 K], a
dehydrogenation enthalpy in the approximate range of 20--50 kJ/mol
\htwo\ would yield desorption pressures/temperatures that are
consistent with the operating conditions of a FC.\cite{schlapbach01}
However, as shown in the last column of Table~\ref{table}, the
calculated $\Delta S_\text{vib}$ are not negligible (up to 21\%) in
comparison to $S_0^{H_2}$, calling into question the assumption
$\Delta S \simeq S_0^{H_2}$ and the guideline $\Delta H$ =
20--50~kJ/mol \htwo.  This suggests that a precise determination of
the pressure-temperature characteristics of a given desorption
reaction requires evaluating the change in Gibbs free energy [$\Delta
G(T)$], accounting explicitly for the effects of temperature and
$\Delta S_\text{vib}$, as done below.

\subsection{Thermodynamic Guidelines}
A key concern when attempting to predict favorable hydrogen storage
reactions is to ensure that the thermodynamically preferred reaction
pathway has been identified.  This is a non-trivial task, and our
experience has shown that intuition alone is not sufficient to
correctly identify realistic reactions involving multicomponent
systems.\cite{siegel07} In this regard, several of the reactions in
Table~\ref{table} (denoted by $*$) are noteworthy as they illustrate
the difficulties that may arise when ``guessing'' at reactions.  For
example, all of the candidate reactions are written as simple,
single-step reactions.  While this may seem reasonable given the
mechanism proposed in Ref.~\onlinecite{vajo05} (Eq.~\ref{destab}) and
its generalization in Eq.~\ref{gen_eqn}, as we discuss below, some of
these reactions should proceed via multiple step pathways, with each
step having thermodynamic properties that are distinct from the
presumed single-step pathway.

We group the examples of how chemical intuition might fail into three
categories, and for each class, give a general guideline describing
the thermodynamic restriction: 

\textit{(1) Reactant mixtures involving
  ``weakly-bound'' compounds:} We refer here to systems where the
enthalpy to decompose one (or more) of the reactant phases is less
than the enthalpy of the proposed destabilized reaction; thus, the
weakly-bound phase(s) will decompose before (i.e., at a temperature
below that which) the destabilized reaction can proceed.  Two examples
of this behavior can be found in Table~\ref{table}.  The first case
pertains to reactions 13--16, which, based on their larger enthalpies
relative to reaction 12, would appear to ``stabilize'' \cabh.  In
reality, \cabh\ will decompose before (with $P_{\text{H}_2} = 1$~bar
at $T$ = 88$^\circ$C) any of the higher temperature reactions 13--16
will occur ($T >$ 110$^\circ$C), indicating that it is impossible to
stabilize a reaction in this manner.  Additional examples of this
scenario occur in reactions 1, 8, 17, and 21, which involve the
metastable \alh\ and \crh\ phases.  In the case of reaction 1, \alh\
will decompose first (yielding Al and $\frac{3}{2}$\htwo), followed by
reaction of Al with \libh\ (reaction 2).  The consequences of this
behavior are significant, since although the intended reaction 1 has
an enthalpy ($\sim$40 kJ/mol \htwo) in the targeted range, in reality
the reaction will consist of two steps, the first of which has an
enthalpy below the targeted range (AlH$_3$ decomposition), while the
second (reaction 2) has an enthalpy above this range.
\textit{Guideline 1: The enthalpy of the proposed destabilized
  reaction must be less than the decomposition enthalpies of the
  individual reactant phases.}

\textit{(2) Unstable combinations of product or reactant phases:}
Reaction 4 illustrates how the seemingly straightforward process of
identifying stable reactant and product phases can become unexpectedly
complex.  Here, the starting mixture of \libh\ and Mg is unstable and
will undergo the exothermic transformation:
\begin{equation}
\label{eq:exolibh}
2 \text{LiBH}_4 + \text{Mg} \rightarrow
\frac{3}{2} \text{LiBH}_4 + \frac{3}{4}\text{MgH}_2 + \frac{1}{4}\text{MgB}_2 + \frac{1}{2}\text{LiH},
\end{equation}
which will consume the available Mg and form \mgh, which will itself
react endothermically with the remaining \libh\ according to reaction
24. The exothermic nature of Eq.~(\ref{eq:exolibh}) can be understood
by noting that the enthalpy of reaction 4 (46.4 kJ/mol \htwo) is lower
than the decomposition enthalpy of MgH$_2$, given by reaction 27 (62.3
kJ/mol \htwo). Therefore, the total energy can be lowered by
transferring hydrogen to the more strongly bound \mgh\ compound.
\textit{Guideline 2: If the proposed reaction involves a reactant that
  can absorb hydrogen (such as an elemental metal), the formation
  enthalpy of the corresponding hydride cannot be greater in magnitude
  than the enthalpy of the destabilized reaction.}

\textit{(3) Lower-energy reaction pathways:} Reaction 3, involving a
4:1 mixture of \libh:\mgh, as well as the related reaction involving a
7:1 stoichiometry, 7\libh\ + \mgh\ $\rightarrow$ MgB$_7$ + 7LiH +
11.5\htwo, were recently suggested in Ref.~\onlinecite{alapati06},
which considered only a single-step mechanism resulting in the
formation of \mgbfour\ and MgB$_7$, respectively. Here we demonstrate
that these reactions will not proceed as suggested there due to the
presence of intermediate stages with lower energies.  In fact, both
hypothetical reactions have larger enthalpies [$\Delta E$ = 69 (4:1)
and 74 (7:1) kJ/mol \htwo\cite{alapati06}] than the 2:1 mixture
(reaction 24), suggesting that, upon increasing temperature, the 4:1
and 7:1 mixtures will follow a pathway whose initial reaction step is
the 2:1 reaction (reaction 24), which will consume all available \mgh.
Subsequent reactions between unreacted \libh\ and newly-formed
\mgbtwo\ will become thermodynamically feasible at temperatures above
that of reaction 24, since their enthalpies exceed 50 kJ/mol \htwo.
[Similar behavior is expected for reactions 9 \& 10, as the 1:1
mixture of \libh:Fe (reaction 9) will initially react in a 1:2 ratio
(reaction 10), which has a lower enthalpy.] \textit{Guideline 3: In
  general, it is not possible to tune the thermodynamics of
  destabilized reactions by adjusting the molar fractions of the
  reactants. There is only one stoichiometry corresponding to a
  single-step reaction with the lowest possible enthalpy; all other
  stoichiometries will release \htwo\ in multi-step reactions, where
  the initial reaction is given by the lowest-enthalpy
  reaction}.\footnote{This discussion assumes that the entropies of all
  competing reaction pathways are similar. Our results in Table I show
  that this is generally not the case; generalization of the above
  guidelines to the free energies is straightforward and will be
  presented elsewhere\cite{siegel07a}.}

\begin{figure}
\centering \includegraphics[width=\linewidth]{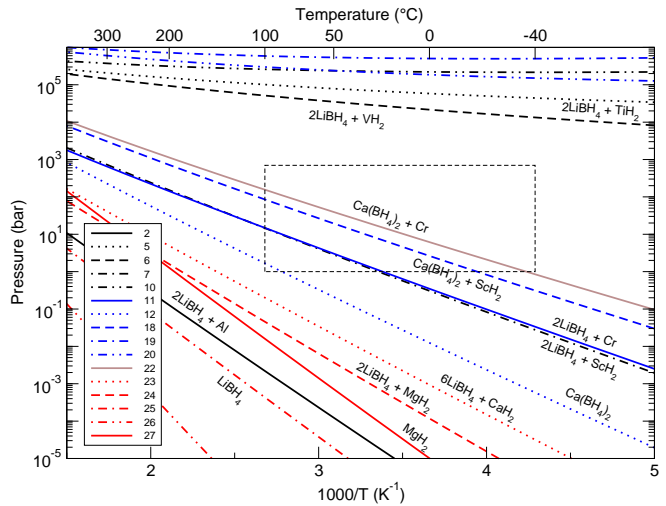}
\caption{(Color online) Calculated van't Hoff plot for reactions
  listed in Table~\ref{table}.  The region within the dashed box
  corresponds to desirable temperatures and pressures for on-board
  hydrogen storage: $P_{\text{H}_2}$ = 1--700~bar, $T$ = -40--100$^\circ$C.}
\label{fig1}
\end{figure}

\subsection{Destabilized Reactions}
In total, the preceding examples reveal that great care must be taken
in predicting hydrogen storage reactions.  Having ruled out the
specious reactions, we now discuss the thermodynamics of the remaining
reactions.  Using the calculated thermodynamic data
(Table~\ref{table}) as input to the van't Hoff equation,
$P_{\text{H}_2} = P_0\exp(-\frac{\Delta G}{RT})$, where $P_0 = 1$~bar,
Fig.~\ref{fig1} plots the equilibrium \htwo\ desorption pressures of
these reactions as a function of temperature.\footnote{We neglect the
  \libh\ structural transition at $T_s \sim
  108^\circ$C\cite{soulie02}, which should reduce the slope of the
  data in Fig.~\ref{fig1} for $T > T_s$.}  Included in the plot is a
rectangle delineating desirable temperature and pressure ranges for
\htwo\ storage: -40--100$^\circ$C, and 1--700~bar.



As expected, our van't Hoff plot confirms that the experimental
reactions having large dehydrogenation enthalpies (reactions 24--27)
yield pressures $P \ll 1$~bar, even at elevated temperatures.  On the
other hand, some of the candidate reactions, for example 5 and 19,
readily evolve \htwo\ at very low temperatures (consistent with their
low enthalpies) and are therefore too weakly bound for practical,
reversible on-board storage.  However, the candidate reactions
involving mixtures with \sch\footnote{It should be noted that the high
  cost of Sc may preclude its use in practical applications.}
(reactions 7\footnote{This reaction was also reported in
  Ref.~\onlinecite{alapati07}} and 18) and Cr (reactions 11 and 22)
desorb \htwo\ in $P$-$T$ regimes that strongly intersect the window of
desirable operating conditions.  These reactions have room-temperature
enthalpies in the range of 27--33 kJ/mol \htwo, relatively high \htwo\ 
densities (5--8.9 wt.\% \htwo\ and 85-100 g \htwo/L), and achieve
$P_{\text{H}_2} = 1$~bar at moderate temperatures ranging from 26 and
$-$38$^\circ$C.  Thus, via a first-principles approach of rapid
screening through a large number of candidate reactions, and the
careful use of thermodynamic considerations to eliminate unstable or
multi-step reactions, we predict here several reactions with
attributes that surpass the state-of-the-art for reversible,
low-temperature storage materials.

\section{Conclusion}
In conclusion, using first-principles free energy calculations we have
demonstrated that further significant destabilization of the
strongly-bound \libh\ and \cabh\ borohydrides is possible, and we
identify several high \htwo-density reactions having thermodynamics
compatible with the operating conditions of mobile \htwo-storage
applications.  Unlike other recent predictions, the proposed reactions
utilize only known compounds with established synthesis routes, and
can therefore be subjected to immediate experimental testing. In
addition, we provide guidance to subsequent efforts aimed at
predicting new \htwo\ storage materials by illustrating common
pitfalls that arise when attempting to ``guess'' at reaction
mechanisms, and by suggesting a set of thermodynamic guidelines to
facilitate more robust predictions.

\begin{acknowledgments}
V.O. thanks the U.S. DOE for financial support under grants
DE-FG02-05ER46253 and DE-FC36-04GO14013.
\end{acknowledgments}


\begin{thebibliography}{32}
\expandafter\ifx\csname natexlab\endcsname\relax\def\natexlab#1{#1}\fi
\expandafter\ifx\csname bibnamefont\endcsname\relax
  \def\bibnamefont#1{#1}\fi
\expandafter\ifx\csname bibfnamefont\endcsname\relax
  \def\bibfnamefont#1{#1}\fi
\expandafter\ifx\csname citenamefont\endcsname\relax
  \def\citenamefont#1{#1}\fi
\expandafter\ifx\csname url\endcsname\relax
  \def\url#1{\texttt{#1}}\fi
\expandafter\ifx\csname urlprefix\endcsname\relax\def\urlprefix{URL }\fi
\providecommand{\bibinfo}[2]{#2}
\providecommand{\eprint}[2][]{\url{#2}}

\bibitem[{\citenamefont{Pinkerton and Wicke}(2004)}]{pinkerton04}
\bibinfo{author}{\bibfnamefont{F.~E.} \bibnamefont{Pinkerton}}
  \bibnamefont{and} \bibinfo{author}{\bibfnamefont{B.~G.} \bibnamefont{Wicke}},
  \bibinfo{journal}{Industrial Physicist} \textbf{\bibinfo{volume}{10}},
  \bibinfo{pages}{20} (\bibinfo{year}{2004}).

\bibitem[{\citenamefont{Crabtree et~al.}(2004)\citenamefont{Crabtree,
  Dresselhaus, and Buchanan}}]{crabtree04}
\bibinfo{author}{\bibfnamefont{G.~W.} \bibnamefont{Crabtree}},
  \bibinfo{author}{\bibfnamefont{M.~S.} \bibnamefont{Dresselhaus}},
  \bibnamefont{and} \bibinfo{author}{\bibfnamefont{M.~V.}
  \bibnamefont{Buchanan}}, \bibinfo{journal}{Physics Today}
  \textbf{\bibinfo{volume}{57}}, \bibinfo{pages}{39} (\bibinfo{year}{2004}).

\bibitem[{\citenamefont{Schlapbach and Z\"{u}ttel}(2001)}]{schlapbach01}
\bibinfo{author}{\bibfnamefont{L.}~\bibnamefont{Schlapbach}} \bibnamefont{and}
  \bibinfo{author}{\bibfnamefont{A.}~\bibnamefont{Z\"{u}ttel}},
  \bibinfo{journal}{Nature} \textbf{\bibinfo{volume}{414}},
  \bibinfo{pages}{353} (\bibinfo{year}{2001}).

\bibitem[{\citenamefont{Sandrock}(1999)}]{sandrock99}
\bibinfo{author}{\bibfnamefont{G.}~\bibnamefont{Sandrock}},
  \bibinfo{journal}{J. Alloys Compd.} \textbf{\bibinfo{volume}{293-295}},
  \bibinfo{pages}{877} (\bibinfo{year}{1999}).

\bibitem[{\citenamefont{Soulie et~al.}(2002)\citenamefont{Soulie, Renaudin,
  Cerny, and Yvon}}]{soulie02}
\bibinfo{author}{\bibfnamefont{J.-P.} \bibnamefont{Soulie}},
  \bibinfo{author}{\bibfnamefont{G.}~\bibnamefont{Renaudin}},
  \bibinfo{author}{\bibfnamefont{R.}~\bibnamefont{Cerny}}, \bibnamefont{and}
  \bibinfo{author}{\bibfnamefont{K.}~\bibnamefont{Yvon}}, \bibinfo{journal}{J.
  Alloys Compd.} \textbf{\bibinfo{volume}{346}}, \bibinfo{pages}{200}
  (\bibinfo{year}{2002}).

\bibitem[{\citenamefont{Zuttel et~al.}(2003)\citenamefont{Zuttel, Rentsch,
  Fischer, Wenger, Sudan, Mauron, and Emmenegger}}]{zuttel03}
\bibinfo{author}{\bibfnamefont{A.}~\bibnamefont{Zuttel}},
  \bibinfo{author}{\bibfnamefont{S.}~\bibnamefont{Rentsch}},
  \bibinfo{author}{\bibfnamefont{P.}~\bibnamefont{Fischer}},
  \bibinfo{author}{\bibfnamefont{P.}~\bibnamefont{Wenger}},
  \bibinfo{author}{\bibfnamefont{P.}~\bibnamefont{Sudan}},
  \bibinfo{author}{\bibfnamefont{P.}~\bibnamefont{Mauron}}, \bibnamefont{and}
  \bibinfo{author}{\bibfnamefont{C.}~\bibnamefont{Emmenegger}},
  \bibinfo{journal}{J. Alloys Compd.} \textbf{\bibinfo{volume}{356-357}},
  \bibinfo{pages}{515} (\bibinfo{year}{2003}).

\bibitem[{\citenamefont{Nakamori et~al.}(2006)\citenamefont{Nakamori, Miwa,
  Ninomiya, Li, Ohba, Towata, Zuttel, and Orimo}}]{nakamori06a}
\bibinfo{author}{\bibfnamefont{Y.}~\bibnamefont{Nakamori}},
  \bibinfo{author}{\bibfnamefont{K.}~\bibnamefont{Miwa}},
  \bibinfo{author}{\bibfnamefont{A.}~\bibnamefont{Ninomiya}},
  \bibinfo{author}{\bibfnamefont{H.}~\bibnamefont{Li}},
  \bibinfo{author}{\bibfnamefont{N.}~\bibnamefont{Ohba}},
  \bibinfo{author}{\bibfnamefont{S.}~\bibnamefont{Towata}},
  \bibinfo{author}{\bibfnamefont{A.}~\bibnamefont{Zuttel}}, \bibnamefont{and}
  \bibinfo{author}{\bibfnamefont{S.}~\bibnamefont{Orimo}},
  \bibinfo{journal}{Phys.\ Rev.\ B} \textbf{\bibinfo{volume}{74}},
  \bibinfo{pages}{045126} (\bibinfo{year}{2006}).

\bibitem[{\citenamefont{{\L}odziana and Vegge}(2004)}]{lodziana04}
\bibinfo{author}{\bibfnamefont{Z.}~\bibnamefont{{\L}odziana}} \bibnamefont{and}
  \bibinfo{author}{\bibfnamefont{T.}~\bibnamefont{Vegge}},
  \bibinfo{journal}{Phys.\ Rev.\ Lett.} \textbf{\bibinfo{volume}{93}},
  \bibinfo{pages}{145501} (\bibinfo{year}{2004}).

\bibitem[{\citenamefont{Miwa et~al.}(2006)\citenamefont{Miwa, Aoki, Noritake,
  Ohba, Nakamori, ichi Towata, Zuttel, and ichi Orimo}}]{miwa06_cabh}
\bibinfo{author}{\bibfnamefont{K.}~\bibnamefont{Miwa}},
  \bibinfo{author}{\bibfnamefont{M.}~\bibnamefont{Aoki}},
  \bibinfo{author}{\bibfnamefont{T.}~\bibnamefont{Noritake}},
  \bibinfo{author}{\bibfnamefont{N.}~\bibnamefont{Ohba}},
  \bibinfo{author}{\bibfnamefont{Y.}~\bibnamefont{Nakamori}},
  \bibinfo{author}{\bibfnamefont{S.}~\bibnamefont{ichi Towata}},
  \bibinfo{author}{\bibfnamefont{A.}~\bibnamefont{Zuttel}}, \bibnamefont{and}
  \bibinfo{author}{\bibfnamefont{S.}~\bibnamefont{ichi Orimo}},
  \bibinfo{journal}{Phys.\ Rev.\ B} \textbf{\bibinfo{volume}{74}},
  \bibinfo{pages}{155122} (\bibinfo{year}{2006}).

\bibitem[{\citenamefont{Reilly and Wiswall}(1968)}]{reilly68}
\bibinfo{author}{\bibfnamefont{J.~J.} \bibnamefont{Reilly}} \bibnamefont{and}
  \bibinfo{author}{\bibfnamefont{R.~H.} \bibnamefont{Wiswall}},
  \bibinfo{journal}{Inorganic Chem.} \textbf{\bibinfo{volume}{7}},
  \bibinfo{pages}{2254} (\bibinfo{year}{1968}).

\bibitem[{\citenamefont{Vajo et~al.}(2005)\citenamefont{Vajo, Skeith, and
  Mertens}}]{vajo05}
\bibinfo{author}{\bibfnamefont{J.~J.} \bibnamefont{Vajo}},
  \bibinfo{author}{\bibfnamefont{S.~L.} \bibnamefont{Skeith}},
  \bibnamefont{and} \bibinfo{author}{\bibfnamefont{F.}~\bibnamefont{Mertens}},
  \bibinfo{journal}{J. Phys.\ Chem.\ B Lett.} \textbf{\bibinfo{volume}{109}},
  \bibinfo{pages}{3719} (\bibinfo{year}{2005}).

\bibitem[{\citenamefont{Manchester}(2000)}]{manchester00}
\bibinfo{editor}{\bibfnamefont{F.~D.} \bibnamefont{Manchester}}, ed.,
  \emph{\bibinfo{title}{Phase Diagrams of Binary Hydrogen Alloys}}
  (\bibinfo{publisher}{ASM}, \bibinfo{address}{Materials Park, OH},
  \bibinfo{year}{2000}).

\bibitem[{\citenamefont{Alapati et~al.}(2006)\citenamefont{Alapati, Johnson,
  and Sholl}}]{alapati06}
\bibinfo{author}{\bibfnamefont{S.~V.} \bibnamefont{Alapati}},
  \bibinfo{author}{\bibfnamefont{J.~K.} \bibnamefont{Johnson}},
  \bibnamefont{and} \bibinfo{author}{\bibfnamefont{D.~S.} \bibnamefont{Sholl}},
  \bibinfo{journal}{J. Phys.\ Chem.\ B} \textbf{\bibinfo{volume}{110}},
  \bibinfo{pages}{8769} (\bibinfo{year}{2006}).

\bibitem[{\citenamefont{Bogdanovi\'{c} and Schwickardi}(1997)}]{bogdanovic97}
\bibinfo{author}{\bibfnamefont{B.}~\bibnamefont{Bogdanovi\'{c}}}
  \bibnamefont{and}
  \bibinfo{author}{\bibfnamefont{M.}~\bibnamefont{Schwickardi}},
  \bibinfo{journal}{J. Alloys Compd.} \textbf{\bibinfo{volume}{253-254}},
  \bibinfo{pages}{1} (\bibinfo{year}{1997}).

\bibitem[{\citenamefont{Deng et~al.}(2004)\citenamefont{Deng, Xu, and
  Goddard}}]{deng04}
\bibinfo{author}{\bibfnamefont{W.-Q.} \bibnamefont{Deng}},
  \bibinfo{author}{\bibfnamefont{X.}~\bibnamefont{Xu}}, \bibnamefont{and}
  \bibinfo{author}{\bibfnamefont{W.~A.} \bibnamefont{Goddard}},
  \bibinfo{journal}{Phys.\ Rev.\ Lett.} \textbf{\bibinfo{volume}{92}},
  \bibinfo{pages}{166103} (\bibinfo{year}{2004}).

\bibitem[{\citenamefont{Zhao et~al.}(2005)\citenamefont{Zhao, Kim, Dillon,
  Heben, and Zhang}}]{zhao05}
\bibinfo{author}{\bibfnamefont{Y.}~\bibnamefont{Zhao}},
  \bibinfo{author}{\bibfnamefont{Y.-H.} \bibnamefont{Kim}},
  \bibinfo{author}{\bibfnamefont{A.~C.} \bibnamefont{Dillon}},
  \bibinfo{author}{\bibfnamefont{M.~J.} \bibnamefont{Heben}}, \bibnamefont{and}
  \bibinfo{author}{\bibfnamefont{S.~B.} \bibnamefont{Zhang}},
  \bibinfo{journal}{Phys.\ Rev.\ Lett.} \textbf{\bibinfo{volume}{94}},
  \bibinfo{pages}{155504} (\bibinfo{year}{2005}).

\bibitem[{\citenamefont{Yildirim and Ciraci}(2005)}]{yildirim05}
\bibinfo{author}{\bibfnamefont{T.}~\bibnamefont{Yildirim}} \bibnamefont{and}
  \bibinfo{author}{\bibfnamefont{S.}~\bibnamefont{Ciraci}},
  \bibinfo{journal}{Phys.\ Rev.\ Lett.} \textbf{\bibinfo{volume}{94}},
  \bibinfo{pages}{175501} (\bibinfo{year}{2005}).

\bibitem[{\citenamefont{Lee et~al.}(2006)\citenamefont{Lee, Choi, and
  Ihm}}]{lee06}
\bibinfo{author}{\bibfnamefont{H.}~\bibnamefont{Lee}},
  \bibinfo{author}{\bibfnamefont{W.~I.} \bibnamefont{Choi}}, \bibnamefont{and}
  \bibinfo{author}{\bibfnamefont{J.}~\bibnamefont{Ihm}},
  \bibinfo{journal}{Phys.\ Rev.\ Lett.} \textbf{\bibinfo{volume}{97}},
  \bibinfo{pages}{056104} (\bibinfo{year}{2006}).

\bibitem[{\citenamefont{Sun et~al.}(2006)\citenamefont{Sun, Jena, Wang, and
  Marquez}}]{sun06}
\bibinfo{author}{\bibfnamefont{Q.}~\bibnamefont{Sun}},
  \bibinfo{author}{\bibfnamefont{P.}~\bibnamefont{Jena}},
  \bibinfo{author}{\bibfnamefont{Q.}~\bibnamefont{Wang}}, \bibnamefont{and}
  \bibinfo{author}{\bibfnamefont{M.}~\bibnamefont{Marquez}},
  \bibinfo{journal}{J. Am. Chem. Soc.} \textbf{\bibinfo{volume}{128}},
  \bibinfo{pages}{9741} (\bibinfo{year}{2006}).

\bibitem[{\citenamefont{Kresse and Furthm{\"{u}}ller}(1996)}]{kresse96}
\bibinfo{author}{\bibfnamefont{G.}~\bibnamefont{Kresse}} \bibnamefont{and}
  \bibinfo{author}{\bibfnamefont{J.}~\bibnamefont{Furthm{\"{u}}ller}},
  \bibinfo{journal}{Phys.\ Rev.\ B} \textbf{\bibinfo{volume}{54}},
  \bibinfo{pages}{11169} (\bibinfo{year}{1996}).

\bibitem[{\citenamefont{Bl{\"{o}}chl}(1994)}]{blochl94a}
\bibinfo{author}{\bibfnamefont{P.~E.} \bibnamefont{Bl{\"{o}}chl}},
  \bibinfo{journal}{Phys.\ Rev.\ B} \textbf{\bibinfo{volume}{50}},
  \bibinfo{pages}{17953} (\bibinfo{year}{1994}).

\bibitem[{\citenamefont{Perdew et~al.}(1992)\citenamefont{Perdew, Chevary,
  Vosko et~al.}}]{perdew92}
\bibinfo{author}{\bibfnamefont{J.~P.} \bibnamefont{Perdew}},
  \bibinfo{author}{\bibfnamefont{J.~A.} \bibnamefont{Chevary}},
  \bibinfo{author}{\bibfnamefont{S.~H.} \bibnamefont{Vosko}},
  \bibnamefont{et~al.}, \bibinfo{journal}{Phys.\ Rev.\ B}
  \textbf{\bibinfo{volume}{46}}, \bibinfo{pages}{6671} (\bibinfo{year}{1992}).

\bibitem[{\citenamefont{Wallace}(1972)}]{wallace72}
\bibinfo{author}{\bibfnamefont{D.~C.} \bibnamefont{Wallace}},
  \emph{\bibinfo{title}{Thermodynamics of Crystals}} (\bibinfo{publisher}{John
  Wiley \& Sons}, \bibinfo{year}{1972}).

\bibitem[{\citenamefont{Wei and Chou}(1992)}]{wei92}
\bibinfo{author}{\bibfnamefont{S.}~\bibnamefont{Wei}} \bibnamefont{and}
  \bibinfo{author}{\bibfnamefont{M.~Y.} \bibnamefont{Chou}},
  \bibinfo{journal}{Phys.\ Rev.\ Lett.} \textbf{\bibinfo{volume}{69}},
  \bibinfo{pages}{2799} (\bibinfo{year}{1992}).

\bibitem[{\citenamefont{Wolverton et~al.}(2004)\citenamefont{Wolverton,
  Ozoli\c{n}\v{s}, and Asta}}]{wolverton04}
\bibinfo{author}{\bibfnamefont{C.}~\bibnamefont{Wolverton}},
  \bibinfo{author}{\bibfnamefont{V.}~\bibnamefont{Ozoli\c{n}\v{s}}},
  \bibnamefont{and} \bibinfo{author}{\bibfnamefont{M.}~\bibnamefont{Asta}},
  \bibinfo{journal}{Phys.\ Rev.\ B} \textbf{\bibinfo{volume}{69}},
  \bibinfo{pages}{144109} (\bibinfo{year}{2004}).

\bibitem[{\citenamefont{Siegel et~al.}(2007{\natexlab{a}})\citenamefont{Siegel,
  Wolverton, and Ozoli\c{n}\v{s}}}]{siegel07}
\bibinfo{author}{\bibfnamefont{D.~J.} \bibnamefont{Siegel}},
  \bibinfo{author}{\bibfnamefont{C.}~\bibnamefont{Wolverton}},
  \bibnamefont{and}
  \bibinfo{author}{\bibfnamefont{V.}~\bibnamefont{Ozoli\c{n}\v{s}}},
  \bibinfo{journal}{Phys.\ Rev.\ B} \textbf{\bibinfo{volume}{75}},
  \bibinfo{pages}{014101} (\bibinfo{year}{2007}{\natexlab{a}}).

\bibitem[{\citenamefont{Wolverton and Ozolins}(2007)}]{wolverton07}
\bibinfo{author}{\bibfnamefont{C.}~\bibnamefont{Wolverton}} \bibnamefont{and}
  \bibinfo{author}{\bibfnamefont{V.}~\bibnamefont{Ozolins}},
  \bibinfo{journal}{Phys.\ Rev.\ B} \textbf{\bibinfo{volume}{75}},
  \bibinfo{pages}{064101} (\bibinfo{year}{2007}).

\bibitem[{\citenamefont{Satyapal et~al.}(2007)\citenamefont{Satyapal, Petrovic,
  Read, Thomas, and Ordaz}}]{doe_targets}
\bibinfo{author}{\bibfnamefont{S.}~\bibnamefont{Satyapal}},
  \bibinfo{author}{\bibfnamefont{J.}~\bibnamefont{Petrovic}},
  \bibinfo{author}{\bibfnamefont{C.}~\bibnamefont{Read}},
  \bibinfo{author}{\bibfnamefont{G.}~\bibnamefont{Thomas}}, \bibnamefont{and}
  \bibinfo{author}{\bibfnamefont{G.}~\bibnamefont{Ordaz}},
  \bibinfo{journal}{Catal.\ Today} \textbf{\bibinfo{volume}{120}},
  \bibinfo{pages}{246} (\bibinfo{year}{2007}).

\bibitem[{\citenamefont{Chase}(1998)}]{janaf98}
\bibinfo{author}{\bibfnamefont{M.~W.} \bibnamefont{Chase}, \bibfnamefont{Jr.}},
  \emph{\bibinfo{title}{NIST-JANAF Thermochemical Tables, 4th Edition}}
  (\bibinfo{publisher}{American Chemical Society}, \bibinfo{year}{1998}).

\bibitem[{asu()}]{asudik_private}
\bibinfo{note}{A. Sudik, private communication.}

\bibitem[{\citenamefont{Siegel et~al.}(2007{\natexlab{b}})\citenamefont{Siegel,
  Ozolins, and Wolverton}}]{siegel07a}
\bibinfo{author}{\bibfnamefont{D.~J.} \bibnamefont{Siegel}},
  \bibinfo{author}{\bibfnamefont{V.}~\bibnamefont{Ozolins}}, \bibnamefont{and}
  \bibinfo{author}{\bibfnamefont{C.}~\bibnamefont{Wolverton}},
  \bibinfo{journal}{Phys.\ Rev.\ B} p. \bibinfo{pages}{in preparation}
  (\bibinfo{year}{2007}{\natexlab{b}}).

\bibitem[{\citenamefont{Alapati et~al.}(2007)\citenamefont{Alapati, Johnson,
  and Sholl}}]{alapati07}
\bibinfo{author}{\bibfnamefont{S.~V.} \bibnamefont{Alapati}},
  \bibinfo{author}{\bibfnamefont{J.~K.} \bibnamefont{Johnson}},
  \bibnamefont{and} \bibinfo{author}{\bibfnamefont{D.~S.} \bibnamefont{Sholl}},
  \bibinfo{journal}{Phys.\ Chem.\ Chem.\ Phys.} \textbf{\bibinfo{volume}{9}},
  \bibinfo{pages}{1438} (\bibinfo{year}{2007}).

\end{thebibliography}

\end{document}